# Symbolic Equation Modeling of Composite Loads: A Kolmogorov-Arnold Network based Learning Approach

Sonam Dorji, *Student Member, IEEE*, Yongkang Sun, *Student Member, IEEE*, Yuchen Zhang, *Member, IEEE*, Ghavameddin Nourbakhsh, *Member, IEEE*, Yateendra Mishra, *Senior Member, IEEE*, Yan Xu, *Senior Member, IEEE*

*Abstract*--With increasing penetration of distributed energy resources installed behind the meter, there is a growing need for adequate modelling of composite loads to enable accurate power system simulation analysis. Existing measurement-based load modeling methods either fit fixed-structure physical models, which limits adaptability to evolving load mixes, or employ flexible machine learning methods which are however "black boxes" and offer limited interpretability. This paper presents a new learning-based load modelling method based on Kolmogorov–Arnold Networks (KANs) towards modelling flexibility and interpretability. By actively learning activation functions on edges, KANs automatically derive free-form symbolic equations that capture nonlinear relationships among measured variables without prior assumptions about load structure. Case studies demonstrate that the proposed approach outperforms other methods in both accuracy and generalization ability, while uniquely representing composite loads into transparent, interpretable mathematical equations.

*Index Terms*--Distributed energy resources, interpretability, Kolmogorov-Arnold networks, load modeling, neural networks

## I. INTRODUCTION

LOAD modeling aims to identify a mathematical representation that accurately reproduces the composite behavior of diverse downstream electrical components [1]. The renewable transition of modern power grids has significantly increased the load complexity, posing challenges in load modeling for reliable power system simulation and analysis.

Advances in grid monitoring infrastructure with phasor measurement units (PMU) have motivated measurement-based load modelling that updates model parameters from on-site electrical measurements. The classic physics-based method relies on an electric load model with a pre-assumed fixed structure, such as ZIP model and composite load model (CLM) [2]. These models, while physically interpretable, often lack flexibility to capture unknown or evolving load compositions.

Learning-based methods leveraging statistical and machine learning techniques have been explored owing to their nonlinear mapping adaptability on unknown relationships [3]. However, they often operate as opaque black boxes, suffering from the lack of interpretability, which limits the physical insights from the models and hinders their integration into existing simulation platforms. Other methods, such as symbolic regression with sparse dictionary learning [4], improve interpretability by selecting and combining basis functions, but its reliance on linear parameter estimation restricts its ability to model highly nonlinear relationships.

Kolmogorov-Arnold Networks (KANs) [5] presents a new machine learning architecture with the potential to bridge this gap between load model flexibility and interpretability. By learning both connection weights and activation functions on edges, KANs automatically derive explicit nonlinear symbolic equations alongside high prediction accuracy. This feature makes KAN distinctive from traditional "black-box" learning architecture with fixed activation functions.

This paper presents the first work introducing KANs into the problem of power system load modeling to the best of authors' knowledge. The proposed KAN-based load modelling approach not only maps the complex nonlinear relationship from disturbance data but also yields transparent, human-readable equations, greatly enhancing interpretability and facilitating integration with existing simulation platforms. A hyperparameter optimization routine is also incorporated to automate a comprehensive KAN training process. Static load modelling validations on a benchmark power system shows that the KAN can reproduce model equations from a single ZIP load and build adaptive load model equations for more complex load compositions, outperforming both multilayer perceptron (MLP) and physics-based methods in terms of accuracy, generalization ability, and interpretability.

## II. LOAD MODELING PROBLEM

The load modeling problem generally refers to accurately representing the electrical behavior of customer loads in power system simulation and analysis, thereby more realistically reflecting their impact on system operation [1]. Mathematically, the purpose of load modelling is to find such an unknown function '$f$' that represents the following load characteristics

$$y = f(x) = f(x_1, x_2, \cdots, x_k) \quad (1)$$

where $y$ represents the dependent electrical variables (typically active power $P$ and reactive power $Q$) responding to the independent variations in $x$; $x$ contains a set of on-site measurable variables as inputs: typically, voltage $V$ and frequency $f$ for static load models, and additional differential

This work was supported by Australian Research Council through its Discovery Early Career Researcher Award (DE220100044).
S. Dorji, Y. Sun, Y. Zhang, G. Nourbakhsh, and Y. Mishra are with the School of Electrical Engineering and Robotics, Queensland University of Technology, Australia (e-mail: s8.dorji@hdr.qut.edu.au, yongkang.sun@hdr.qut.edu.au, yuchen1.zhang@qut.edu.au, g.nourbakhsh@qut.edu.au, yateendra.mishra@qut.edu.au).
Yan Xu is with the School of Electrical and Electronic Engineering, Nanyang Technological University, Singapore (email: xuyan@ntu.edu.sg).

terms for dynamic load models.

Reliable dynamic simulation, stability assessment, voltage regulation and operational planning all depend on load models that remain valid under time-varying, evolving, and heterogeneous load compositions [4]. Existing measurement-based methods either fix the load structure and only tune parameters, which limit the adaptability, or use flexible machine learning-based methods that are non-interpretable. What is needed is a data-driven identification of $f$ that adapts to unknown nonlinear behavior while yielding explicit, free-form, and inspectable presentation. The KAN-based method is introduced to achieve this.

## III. PROPOSED LOAD MODELING METHOD

### A. Kolmogorov-Arnold Representation Theorem

The Kolmogorov-Arnold representation theorem [6] states that any multivariate continuous function $f$ defined on a bounded domain can be represented by a finite composition of continuous univariate and binary addition operations. For any smooth function $f : [0,1]^k \to R$ , there exists a series of continuous univariate functions $\phi_{p,q}$ and $\psi_q$ such that:

$$f(x) = f(x_1, x_2, \cdots, x_k) = \sum_{q=1}^{2k+1} \Phi_q \left( \sum_{p=1}^{k} \phi_{p,q}(x_p) \right) \qquad (2)$$

where $x_p = (x_1, x_2, \cdots, x_k)$ , $\phi_{p,q} : [0,1] \to R$ and $\Phi_q : R \to R$ are continuous functions. This theorem demonstrates that learning a high-dimensional function can be reduced to learning a polynomial combination of univariate functions.

### B. Kolmogorov-Arnold Networks

The activation value of the $(l+1, j)$ neurons is computed as the sum of all incoming post-activations [5]:

$$x_{l+1,j} = \sum_{i=1}^{n_l} \tilde{x}_{l,j,i} = \sum_{i=1}^{n_l} \phi_{l,j,i}(x_{l,i}) \qquad (3)$$

where $n_l$ denotes the number of neurons in the $l^{th}$ layer, $\phi_{l,j,i}(.)$ is the activation function connecting the $i^{th}$ neuron in the $l^{th}$ layer and the $j^{th}$ neuron in the $(l+1)^{th}$ layer, which is a trainable; and $\tilde{x}_{l,j,i}$ refers to the post-activation value.

The general composition of a KAN network with $L$ layers for a given input vector $x \in R^{n_0}$ can be written as:

$$KAN(x) = (\Phi_{L-1} \cdot \Phi_{L-2} \cdots \cdot \Phi_1 \cdot \Phi_0) x \qquad (4)$$

where $\Phi$ denotes the matrix formed by concatenating all the activation functions assigned to every edge in a single layer.

Equation (4) defines a layered structure where each layer transforms its input via a matrix of univariate functions. Fig. 1 illustrates a typical three-layer KAN structure, where input vector passes sequentially through layers, each containing trainable univariate activation functions learned from the previous layer's output.

Although the KAN formulation in (3) looks straightforward but achieving optimal performance through its optimization is a challenge. However, [5] incorporates the residual activation functions, where the activation function $\phi(x)$ is the sum of basis function $b(x)$ and spline function as below:

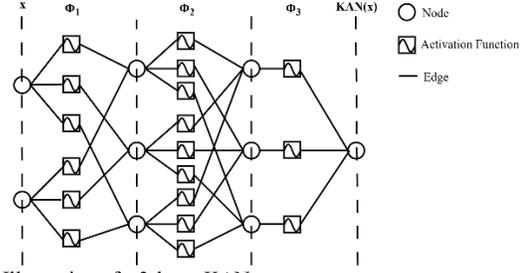

Fig.1. Illustration of a 3-layer KAN structure.

$$\phi(x) = w_b b(x) + w_s spline(x) \qquad (5)$$

where $b(x)$ is defined as $\dfrac{x}{(1+e^{-x})}$ , the spline function is parameterized as a linear combination of B-splines such that, $spline(x) = \sum_i c_i B_i(x)$ , where $c_i s$ are trainable parameters.

### C. KAN-based Load Modeling

The proposed KAN-based load modeling framework, illustrated in Fig. 2, uses input-output data collected from on-site measurements during and after a disturbance at the point of connection of load. The data is preprocessed using Z-score normalization to ensure consistent data scaling. Bayesian Optimization (BO) were employed for hyperparameter tuning to determine optimal values of KAN hyperparameters. Upon obtaining optimal parameters, the KAN model was constructed and trained using the Limited-memory Broyden-Fletcher-Goldfarb-Shanno (LBFGS) optimization algorithm.

Model performances were evaluated using standard error metrics. Once acceptable validation errors were achieved, symbolic equations were extracted from the fully trained model. These equations were then postprocessed to improve readability. The resulting symbolic expressions can be seamlessly integrated into power system simulation platform.

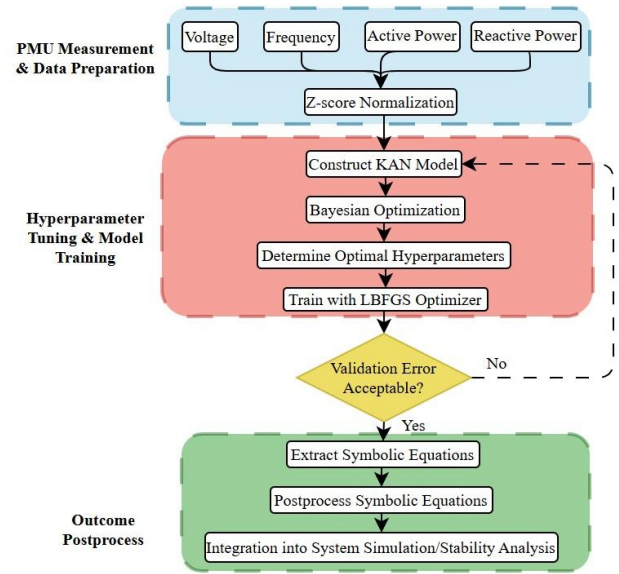

Fig. 2. The process of training KANs-based load model.

## IV. CASE STUDY

### A. Simulation Setup and Data Generation

The simulations to generate load response data are conducted on a two-area Kurdur system using PowerWorld Simulator, with a focus on static load modelling at bus 9. Three fault scenarios are considered: a three-phase fault applied on bus 3, bus 10, and bus 11. In each scenario, the fault is initiated at 1.0 second and cleared at 1.2 seconds. The total simulation time is 10 seconds, and phasor data are recorded at 120 samples per second. The disturbance data from the bus 10 fault serves as the training data for load models, while the data from the bus 3 and bus 11 faults are used as validation and testing data, respectively. The data is prepared to assess the generalization performance of different static load models.

### B. Testing Results on a Single Load Component

In the first case, a pure ZIP load model is connected at bus 9. KAN is trained using the corresponding disturbance data, without being given the true model structure. The symbolic load equations modelled by KAN are displayed in (6) and (7). Despite the lack of prior knowledge, KAN successfully reproduces the polynomial expression of the true ZIP load model from its free-form training. This demonstrates KAN's capability to infer underlying nonlinear voltage-dependent relationships and its alignment with physical load behavior through symbolic equation modeling.

$$P = 32.594V^2 + 464.008V + 1663.861 \qquad (6)$$

$$Q = 4.3619V^2 + 62.5527V + 224.2599 \qquad (7)$$

### C. Testing Results on Complex Load Composition

The second case is conducted to evaluate KAN's ability to extract a static load model from a more realistic and heterogeneous load composition. A composite load model (CLM) and a PV generation unit are connected at bus 9 to represent the scenario with inverter-based distributed generation (DG). KAN's performance is compared with a black-box MLP, a ZIP and an exponential load model with coefficients fitted via least squares. Testing results in Table I employ normalized error metrics as performance metrics for both active and reactive power. The KAN-based load model consistently outperforms other models across all metrics, indicating superior load modeling accuracy and generalization. Figure 3 visually illustrates actual versus predicted active power across four methods. The proposed KAN-based approach captures the deep initial drop and oscillatory recovery with the smallest transient deviation and minimal steady-state bias, further confirming its improved fidelity of load modelling.

TABLE I
Performance Comparison for Active and Reactive Power Estimation
on CLM+DG using Different Load Modelling Methods

| Model | Active Power | | | Reactive Power | | |
|---|---|---|---|---|---|---|
| | MSE | RMSE | MAE | MSE | RMSE | MAE |
| KAN (Proposed) | 0.0352 | 0.1878 | 0.0698 | 0.0426 | 0.2064 | 0.0722 |
| MLP | 0.1136 | 0.3370 | 0.3080 | 0.2191 | 0.4681 | 0.2316 |
| ZIP | 0.4901 | 0.7001 | 0.4753 | 0.4931 | 0.7022 | 0.3643 |
| Exponential | 0.6662 | 0.8162 | 0.5941 | 3.675 | 1.9174 | 1.7841 |

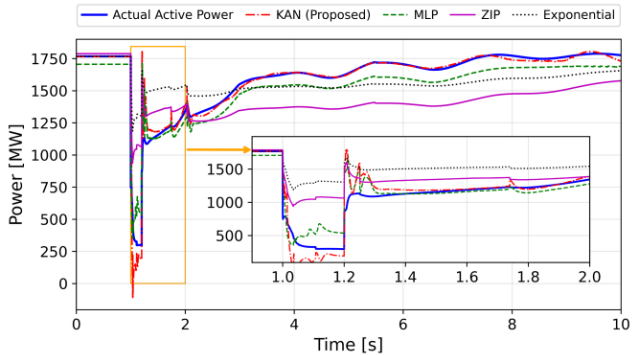

Fig. 3. Actual and predicted active power response.

Additionally, the explicit symbolic expressions learned by KAN for active and reactive power in (8) and (9) capture the dependencies on voltage $V$ and frequency $f$ in unique forms. These expressions more accurately reflect voltage and frequency-dependent characteristics using a static load model.

$$P = -4.9957f + 15.8267e^{1.4334V} + 1561.8429 \qquad (8)$$

$$Q = -36.5522f - 117.7327(-V - 0.4925)^2 + 505.9274 \qquad (9)$$

## V. CONCLUSION

This paper introduces a novel KAN-based approach for both flexible and interpretable modelling of composite loads through automatic symbolic equation extraction. Simulation studies demonstrate that the KAN approach outperforms machine learning models and physics models in static load modeling accuracy, robustness, and generalization. By combining strong modelling capability with interpretable symbolic equations to represent underlying nonlinear relationships, KANs provide a compelling foundation for next-generation intelligent and interpretable load models in modern power systems.

Future works will focus on extending KAN to dynamic load modelling by incorporating physics-informed structures, enhancing the post-processing of symbolic equations for better alignment with known physical relationships and easier interpretation, and exploring more efficient training algorithms to improve computational scalability.